\documentstyle[amssymb,aps,12pt]{revtex}
\begin{document}
\draft
\author{Sergio De Filippo\cite{byline}}
\address{Dipartimento di Scienze Fisiche, Universit\`a di Salerno\\
Via Allende I-84081 Baronissi (SA) ITALY\\
and \\
Unit\`{a} INFM Salerno}
\date{\today}
\title{Gravitationally induced transition to classical behavior above $10^{10}$
proton masses.}
\maketitle

\begin{abstract}
The localization length for the center of mass motion of a matter lump,
induced by gravitation, is obtained, without using any phenomenological
constants. Its dependence from mass and volume is consistent both with
unitary evolution of microscopic particles and with the classical behavior
of macroscopic bodies required to account for wave function collapse in
quantum measurements, the transition between the two regimes being rather
sharp. The gravitational interaction of nonrelativistic matter is modelled
by a Yukawa Hamiltonian with vanishing pion mass, no gravitational
background is needed and the only hypothesis consists in assuming
unentanglement between matter and the Yukawa field.
\end{abstract}


Reconciling gravitation and microphysics\cite{dirac1} has been one of the
fundamental open theoretical problems since the birth of Quantum Mechanics
(QM)\cite{dirac2,von}, even more challenging after the renormalization of
electrodynamics by Feynman, Tomonaga, and Schwinger\cite{schwinger}. On the
other hand, according to several authors\cite{bell,ghirardi,pearle,percival}%
, the conventional interpretation of QM is not completely satisfactory due
to the dualistic description it gives for measurement processes and for time
evolution of isolated microscopic systems.

A possible link between these two issues has been suggested on several
grounds\cite{karolyhazy,diosi,penrose}. However, if we accept that the
solution of the measurement problem in QM, which is central to its
experimental relevance, should be derived from a consistent unified theory
including QM and gravity, which on the other hand, apart from possible
astrophysical and cosmological implications, is hardly expected to have any
direct experimental evidence, we run the risk of abandoning a well
established and rewarding scientific paradigm. We may be lured to admit
that, in order to build a consistent atomic, molecular and condensed matter
physics, we need a detailed knowledge of the physics at the Planck scale,
while one would expect that the remnants of such faraway lengths and
energies in everyday physics could be collected in some constants, which, in
a phenomenological theory at low energies, should be fitted experimentally.

Within a phenomenological approach some authors\cite
{ghirardi,pearle,percival} introduced classical stochastic external elements
in order to produce the wave function collapse. In that context, in order to
fit the experimental evidence for unitary evolution of sufficiently small
systems and for localization of macroscopic bodies, a localization length
and a time constant are usually fixed as phenomenological constants.

As to the link between collapse models and gravitation, two alternative
attitudes are conceivable.

One can just use gravitation as little more than an alibi, supported for
instance by the consideration that internal excitations of atomic systems
induced by the interaction with an external stochastic field are minimized,
and made largely compatible with the existing experimental upper bounds, if
the coupling is proportional to the particle mass\cite{squires}. This is
intuitively evident, since in such a case for long enough wavelengths of the
stochastic field the main coupling is to the center of mass of the atomic
system.

An alternative option consists in starting from gravitation in order to look
for quantitative characterizations of the localization process\cite
{diosi,penrose,anandan,power}. From this viewpoint, even if one maintains
that the Copenhagen formulation of QM, at most supplemented by a multiworld
interpretation\cite{everett,dewitt}, is essentially complete, and appeals to
the environment-induced decoherence program only\cite{zurek,unruh}, to
account for the localization of macroscopic bodies, one should at least
consider the gravitational field as included in the environment.

In this paper we choose the latter option, without addressing the search for
a theory of everything\cite{weinberg,green}, but on the contrary limiting
our ambitions and looking for quantitative results on gravitationally
induced localization of nonrelativistic matter. To be specific we are going
to focus on two fundamental questions:

1) Is the gravitational field vacuum able to localize matter lumps without
the need for a gravitational background of cosmological origin?

2) Is the transition to classical behavior, induced by localization, smooth
or sharp?

As a result we will find an expression for the mass and volume dependent
localization length in the absence of such a background and corresponding to
a rather sharp transition.

A peculiarity of our result consists, at variance with current collapse
models, in a substantial ineffectiveness of the gravitational vacuum in
localizing microscopic particles independently from how long we wait. This
avoids the need of fixing a large time constant like other authors\cite
{ghirardi,pearle,percival} are forced to do, to limit unwanted localization
of microscopic particles. On the contrary we get a rather sharp transition
to an extremely effective localization of macroscopic lumps of condensed
matter at around $10^{10}$ proton masses, this conferring a classical
character to the center of mass motion of macroscopic bodies.

Our main hypothesis stems from the need of reconciling two apparently
conflicting requirements. On one side the assumption of a wave function
collapse induced by the gravitational field requires its classical
character. On the other hand our intention of treating it dynamically and
not just as an external noise forces us, for consistency with the quantum
nature of matter, to quantize it. On different grounds such paradoxical
conditions are reflected in Feynman's words ''...maybe nature is trying to
tell us something here, maybe we should not try to quantize gravity''\cite
{feynman} within his lectures on gravitation. Our proposed way out of this
puzzle consists in assuming that the hidden physics of a consistent unified
theory is continuously collapsing the entanglement between the matter and
the pion field.

To be specific consider nonrelativistic particles of mass $m$ whose
interaction Hamiltonian with a scalar gravitational potential $\hat{\phi}%
(x)\;$is given by 
\begin{equation}
H_{I}\left[ \hat{n}({\bf k})\right] \equiv m\sqrt{4\pi c\hslash G}%
\int\limits_{R^{3}}d{\bf k}\frac{\hat{n}({\bf k})}{\sqrt{k}}\left[ \hat{a}(%
{\bf k)}+a^{\dagger }(-{\bf k)}\right] =mc\sqrt{4\pi G}\int\limits_{R^{3}}d%
{\bf x}\hat{\psi}^{\dagger }({\bf x})\hat{\psi}({\bf x})\hat{\phi}({\bf x}),
\label{HI}
\end{equation}
where $\hat{n}({\bf k})$ is the Fourier transform of the product $\hat{\psi}%
^{\dagger }({\bf x})\hat{\psi}({\bf x})$ (where sum over spin indices is
implied) of the matter field creation and annihilation operators, $\hat{a}%
^{\dagger }({\bf k)}$ and $\hat{a}({\bf k)}$ are the creation and
annihilation operators of the corresponding mode of the scalar potential, $%
\hslash $ and $G$ the Planck and the gravitational constants. The total
Hamiltonian is taken to be the sum of $H_{I}$, of the matter Hamiltonian,
and of 
\begin{equation}
H_{G}=c\hslash \int\limits_{R^{3}}d{\bf k}k\hat{a}^{\dagger }({\bf k)}\hat{a}%
({\bf k),}
\end{equation}
this making our model the analog, for vanishing pion mass, of the Yukawa low
energy theory of nuclear interactions\cite{yukawa}, by which the scalar
particles associated with the $\hat{\phi}$ field will be referred to as
(gravitational) pions.

Of course one could obtain this nonrelativistic model starting from general
relativity, keeping just the conformal excitations of a flat vacuum \cite
{rosales} and then linearizing and quantizing the corresponding Hamiltonian.
We prefer to keep the analysis completely independent from Einstein
equations of the gravitational field, which presumably are only the large
scale manifestation of a fundamental theory that may be well out of reach.
Furthermore, whereas we are not able to quantize the Einstein version of the
gravitational field, this quantum model reproduces the $1/r$ interaction
potential as the limit of the Yukawa potential for vanishing pion mass and
can then be considered as a low energy effective model for gravitation. As a
consequence we consider this model more likely of having a direct quantum
relevance.

While our procedure reminds nonrelativistic electrodynamics, a substantial
difference is to be kept in mind. For electrodynamics a quantization of the
Coulomb field alone is not viable, since a scalar Yukawa-like theory can
only produce attraction between like particles. Then, when in
nonrelativistic classical electrodynamics one fixes the Coulomb gauge, the
Coulomb interaction is separated from the radiation field and inserted into
the matter Hamiltonian so that, if one confines his attention to Coulomb
interaction, only the matter degrees of freedom are left. On the contrary
here pion degrees of freedom are explicitly responsible for the $1/r$ law.

To be specific, if one for simplicity replaces $\hat{n}({\bf k})$ in Eq. (%
\ref{HI}) with the expression, $1+\exp [i{\bf k}\cdot {\bf r}]$,
corresponding to two classical point sources respectively in the origin and
in the point ${\bf r}$, one finds that, while of course the ground state
energy of the $\hat{\phi}$ field in the presence of these sources diverges
due to their singular character, its gradient is well defined and gives the
expected $1/r^{2}$ force law. If the two masses are left free to move, then
the kinetic energy they acquire in falling towards the common center of mass
is balanced by the decrease of the ground state energy of the coherent pion
cloud.

This simple exercise gives a relevant clue to answer the first
aforementioned fundamental question. At first sight the answer is in the
negative since obviously some energy is needed to localize matter and it is
precisely this energetic effect to be considered one of the possible
experimental evidences of fundamental localization processes\cite{squires}.
On the other hand, if one considers that the $\hat{\phi}$ vacuum is not the
ground state in the presence of matter, one has to assume that the coherent 
{\it pion cloud} corresponding to the $\hat{\phi}$ ground state depends on
the matter state. Then the localization process is accompanied by the
rearrangement of the $\hat{\phi}$\ ground state, which is here too the
source of the incremental particle kinetic energy.

Let's begin by evaluating the gravitational ground state energy in the
presence of the $N$-particle matter state $\left| \Psi \right\rangle $
corresponding for computational simplicity to an isotropic Gaussian mass
density of dispersion $\lambda $: 
\begin{equation}
\rho ({\bf x)=}m\left\langle \Psi \right| \hat{\psi}^{\dagger }({\bf x})\hat{%
\psi}({\bf x})\left| \Psi \right\rangle =\frac{Nme^{-\frac{x^{2}}{2\lambda
^{2}}}}{(2\pi )^{3/2}\lambda ^{3}}\Rightarrow \;n_{\lambda }({\bf k})\equiv 
\frac{1}{\left( 2\pi \right) ^{3/2}}\int d{\bf x}\frac{\rho ({\bf x)}}{m}%
e^{-i{\bf k}\cdot {\bf x}}=\frac{Ne^{-\frac{\lambda ^{2}k^{2}}{2}}}{(2\pi
)^{3/2}}.
\end{equation}
In order to do that we observe that the product state $\left| \Psi
\right\rangle \otimes \left| n_{\lambda }\right\rangle _{\phi }$ is the
dressed matter state with the pion cloud in its ground state, namely it is
the minimum energy state for $H_{G}+H_{I}[n_{\lambda }({\bf k})]$ with a
fixed particle factor $\left| \Psi \right\rangle ,$ if $\left| n_{\lambda
}\right\rangle _{\phi }$ is the coherent state 
\begin{equation}
\left| n_{\lambda }\right\rangle _{\phi }=\exp \left[ -m^{2}\frac{4\pi G}{%
c\hslash }\int d{\bf k}k^{-3}n_{\lambda }({\bf k})^{2}\right] \exp \left[ m%
\sqrt{\frac{4\pi G}{c\hslash }}\int d{\bf k}k^{-3/2}n_{\lambda }({\bf k})%
\hat{a}^{\dagger }({\bf k)}\right] \left| 0\right\rangle _{\phi },
\end{equation}
where $\left| 0\right\rangle _{\phi }$ denotes the pion vacuum. This
immediately follows from the expression 
\begin{equation}
H_{G}+H_{I}[n_{\lambda }({\bf k})]=c\hslash \int\limits_{R^{3}}d{\bf k}k\hat{%
b}^{\dagger }({\bf k)}\hat{b}({\bf k)}+E_{0}(\lambda ),
\end{equation}
where 
\begin{equation}
\hat{b}({\bf k})\equiv \hat{a}({\bf k})+m\sqrt{\frac{4\pi G}{c\hslash }}%
\frac{n_{\lambda }({\bf k})}{k^{3/2}}
\end{equation}
and the pion ground state energy is given by 
\begin{equation}
E_{0}(\lambda )=-4\pi m^{2}G\int \frac{d{\bf k}}{k^{2}}\left| n_{\lambda }(%
{\bf k})\right| ^{2}=-4\pi G\frac{M^{2}}{\lambda }.
\end{equation}

Assume now that we have an isolated matter lump and that its wave function
is the product of a wave function of the center of mass ${\bf x}$ 
\begin{equation}
_{CM}\Psi _{\lambda ^{\prime }}({\bf x)=}\frac{\exp \left[ -\frac{x^{2}}{%
4\lambda ^{^{\prime }2}}\right] }{(2\pi )^{3/4}\lambda ^{\prime 3/2}}
\end{equation}
with Gaussian squared modulus of dispersion $\lambda ^{\prime }$, and the
wave function of some stationary inner state, whose particle density, for
computational simplicity, is approximated by a Gaussian of dispersion $%
\lambda _{0}$. Then, ignoring correlations, the corresponding mass density
has dispersion $\lambda $=$\sqrt{\lambda _{0}^{2}+\lambda ^{\prime 2}}$.

The total energy of the system including the matter lump and its pion cloud
is the sum of the pion ground state energy and the matter kinetic energy 
\begin{equation}
E_{T}(\lambda ^{\prime })=E_{0}(\sqrt{\lambda _{0}^{2}+\lambda ^{\prime 2}})-%
\frac{\hslash ^{2}}{2M}\int\limits_{R^{3}}\;_{CM}\Psi _{\lambda ^{\prime }}(%
{\bf x)}\nabla _{CM}^{2}\Psi _{\lambda ^{\prime }}({\bf x)}d{\bf x=-}\frac{%
4\pi GM^{2}}{\sqrt{\lambda _{0}^{2}+\lambda ^{\prime 2}}}+\frac{3\hslash ^{2}%
}{8M\lambda ^{\prime 2}},\;M\equiv Nm.
\end{equation}

The $\lambda ^{\prime }$ value for which $E_{T}(\lambda ^{\prime })$ attains
its minimum is the maximum localization length, since less localized states
are unstable with respect to conversion of pion field energy into kinetic
energy, just like the two point masses mentioned above.

The crucial equation is then 
\begin{equation}
\frac{d}{d\lambda ^{\prime }}E_{T}(\lambda ^{\prime })=0\Rightarrow \left(
\lambda _{0}^{2}+\lambda ^{\prime 2}\right) ^{3}=\frac{64}{9}\pi
^{2}G^{2}\hslash ^{-4}M^{6}\lambda ^{\prime 8},  \label{maineq}
\end{equation}
and, to be specific, we can chose a typical mass density value corresponding
to condensed matter $M/\lambda _{0}^{3}=10^{24}m_{p}/cm^{3}$, where $m_{p}$
denotes the proton mass, so that Eq. (\ref{maineq}) becomes 
\begin{equation}
\left( 10^{-16}\mu ^{2/3}cm^{2}+\lambda ^{\prime 2}\right) ^{3}=\frac{64}{9}%
\pi ^{2}G^{2}\hslash ^{-4}m_{p}^{6}\mu ^{6}\lambda ^{\prime 8},\;\mu \equiv
M/m_{p}.
\end{equation}
This equation gives 
\begin{equation}
\lambda ^{\prime }\simeq 4.23\cdot 10^{23}\mu ^{-3}cm\;\;if\;\;\mu \lesssim
10^{9},\;\;\;\;\;\;\;\;\;\;\lambda ^{\prime }\simeq .81\mu
^{-1/2}cm\,\;\;if\;\;\mu \gtrsim 10^{10},
\end{equation}
which shows that around $M\approx 10^{10}m_{p}$ there is a rather sharp
transition to classical behavior.

This result implies that the localization due to the gravitational vacuum is
microscopically irrelevant even at molecular level, while it accounts for
localization of macroscopic bodies and in particular for the collapse of the
pointer states in quantum measurement theory. Apart from accounting for this
elusive feature of QM, this localization length could have relevant physical
consequences at mesoscopic level, where a localization length $\lambda
^{\prime }\approx 10^{-5}cm$ is comparable with the dimensions of a
corresponding lump of condensed matter with $m\sim 10^{10}m_{p}$.

We can conclude that, within the present nonrelativistic setting and with
the only hypothesis of unentanglement between matter and the pion field, it
is proven that gravitation accounts for wave function collapse without
assuming a gravitational background of cosmological origin. Of course this
does not rule out that the effects of such a background, which may depend on
the particular space-time region, should be added to the more fundamental
ones proposed here. Furthermore one should remark that, while environment
induced decoherence, without including gravitational effects, if not
properly controlled, may overshadow gravitational localization, it can only
account for the entanglement between measured systems and pointer states,
but not for the final collapse of the entangled state.

Finally one should observe that the expression for $E_{0}(\lambda )$ could
have been obtained more naively even without quantizing the pion field. We
chose to put the {\it classical agent} outside the considered model. Of
course the final result only shifts the ''Heisenberg cut''\cite{bell2} to a
consistent unified theory outside the scope of the present letter. However
we are encouraged in this attempt by Feynman's words ''People say to me,
`Are you looking for the ultimate laws of physics?' No, I'm not.''\cite
{feynman2}

Acknowledgments - Financial support from M.U.R.S.T., Italy and I.N.F.M.,
Salerno is acknowledged


\begin{references}
\bibitem[*]{byline}  Phone: +39 89 965229; FAX: +39 89 965275; electronic
address: defilippo@sa.infn.it

\bibitem{dirac1}  P.M.A. Dirac, {\it Lectures on Quantum Mechanics}. Belfer
Graduate School of Science. Yeshiva University, NY (1964).

\bibitem{dirac2}  P.M.A. Dirac, {\it The Principles of Quantum Mechanics}.
Clarendon Press, Oxford (1947).

\bibitem{von}  J. von Neumann, Mathematische Grundlagen der Quantenmechanik
(Springer, Berlin 1932)

\bibitem{schwinger}  {\it Selected Papers on Quantum Electrodynamics}, J.
Schwinger ed., Dover Publications, Inc., NY (1958).

\bibitem{bell}  J. S. Bell, in {\it Sixty-Two Years of Uncertainty}, A. I.
Miller ed., Plenum, NY (1990), p.17; in {\it Themes in Contemporary Physics
II, Essays in Honor of Julian Schwinger's 70th Birthday}, S. Deser and R. J.
Finkelstein eds., World Scientific, Singapore (1989), p. 1.

\bibitem{ghirardi}  G. C. Ghirardi, A. Rimini and T. Weber, Phys. Rev. D 
{\bf 34}, 470 (1986); Phys. Rev.D {\bf 36}, 3287 (1987); Found. Phys. {\bf 18%
}, 1, (1988).

\bibitem{pearle}  P. Pearle, Physical Review A {\bf 39}, 2277 (1989); G. C.
Ghirardi, P. Pearle and A. Rimini, Physical Review A {\bf 42}, 78 (1990).

\bibitem{percival}  I.C. Percival, Proc. R. Soc. Lond. A {\bf 451}, 503
(1995).

\bibitem{karolyhazy}  F. Karolyhazy, Nuo. Cim. A {\bf 42}, 390 (1966); F.
Karolyhazy, A. Frenkel and B. Lukacs B., in . {\it Quantum Concepts in Space
and Time, }R. Penrose and C.J. Isham, eds. Clarenden, Oxford Science
Publications,Oxford (1986).

\bibitem{diosi}  L. Diosi, Phys. Rev. A {\bf 40}, 1165 (1989).

\bibitem{penrose}  R. Penrose, Gen. Rel. Grav. {\bf 28}, 581 (1996).

\bibitem{squires}  P. Pearle, E. Squires, Phys. Rev. Lett. {\bf 73}\ (1994)
1; P. Pearle, J. Ring, J. I. Collar and T. Frank, Found. Phys. {\bf 29}, 465
(1999).

\bibitem{everett}  H. Everett, Revs. Mod. Phys {\bf 29}, 454 (1957).

\bibitem{dewitt}  B. S. DeWitt, in {\it The Many Worlds Interpretation of
Quantum Mechanics}, B. S. Dewitt and N. Graham eds. (Princeton University
Press, Princeton 1973).

\bibitem{anandan}  J. Anandan, Found. Phys. {\bf 29}, 333 (1999).

\bibitem{power}  W.L. Power, I. C. Percival, Proc. Roy. Soc. Lond. {\bf A456}%
, 955 (2000).

\bibitem{zurek}  W.H. Zurek, Phys. Today {\bf 44}, 36 (1991).

\bibitem{unruh}  W.G. Unruh, Phys. Rev. A {\bf 51}, 992 (1995).

\bibitem{weinberg}  S. Weinberg, {\it Dreams of a Final Theory} (Pantheon,
New York, 1992).

\bibitem{green}  B. Green, {\it The Elegant Universe }(Norton, NY, 1999)

\bibitem{feynman}  R.P. Feynman, F.B. Moringo and W.G. Wagner, 1962-3
Lectures on Gravitation. California Institute of Technology (1973), p.12.

\bibitem{yukawa}  H. Yukawa, Proc. Phys.-Math. Soc. Japan {\bf 17}, 48
(1935).

\bibitem{rosales}  J.L. Rosales, J.L. Sanchez-Gomez, Phys. Lett. A {\bf 199}%
, 320 (1995).

\bibitem{bell2}  J.S. Bell {\it Speakeble and Unspeakeble in Quantum
Mechanics} (Cambridge Univ. Press, 1987).

\bibitem{feynman2}  {\it No Ordinary Genius - The Illustrated Richard
Feynman, }Christopher Sykes, W.W.Norton and Company, (1994), p251.
\end{references}
\end{document}